\documentclass[review]{elsarticle}

\usepackage[T1]{fontenc}
\usepackage{geometry}
\usepackage{lineno,hyperref}
\usepackage{graphicx}
\usepackage{appendix}
\usepackage{subfig}
\usepackage{float}
\usepackage{amsmath}
\usepackage{bm}
\usepackage{amssymb}
\usepackage{contour,color,CJK}
\usepackage{lipsum}
\usepackage{booktabs}
\usepackage{epstopdf}
\usepackage{pgfplots}
\usetikzlibrary{patterns}
\usetikzlibrary{spy}
\usepackage{bm}
  \usetikzlibrary{plotmarks}
  \usepgfplotslibrary{patchplots}
\usepackage{grffile}
\pgfplotsset{plot coordinates/math parser=false}
  \newlength\figureheight
  \newlength\figurewidth
\newsavebox{\measurebox}
\usepackage{mathtools}
\DeclarePairedDelimiter{\abs}{\lvert}{\rvert}
\allowdisplaybreaks
\usepackage{etoolbox,ragged2e,siunitx}
\usepackage{nomencl}
\usepackage{xstring}
\usepackage{xpatch}
\usepackage{multicol}
\usepackage{multirow}
\usepackage{array}
\usepackage{framed} 
\usepackage{multicol} 

\renewcommand*\nompreamble{\begin{multicols}{2}}
\renewcommand*\nompostamble{\end{multicols}}

\makenomenclature
\modulolinenumbers[5]

\biboptions{numbers,sort&compress}
\let\oldbibliography\thebibliography
\renewcommand{\thebibliography}[1]{%
  \oldbibliography{#1}
  \setlength{\itemsep}{0pt}}

\begin{document}

\begin{frontmatter}

\title{Minimizing finite viscosity enhances relative kinetic energy absorption in bistable mechanical metamaterials but only with sufficiently fine discretization: a nonlinear dynamical size effect}

\author{Haning Xiu, Ryan Fancher, Ian Frankel}
\address{Department of Mechanical and Aerospace Engineering, University of California, San Diego, La Jolla, CA, US}
\author{Patrick Ziemke}
\address{Department of Mechanical Engineering University of California, Santa Barbara, Santa Barbara, CA, US}
\address{Materials Department, University of California, Santa Barbara, Santa Barbara, CA, US}
\author{M\"{u}ge Fermen-Coker}
\address{DEVCOM Army Research Laboratory, Aberdeen Proving Ground, MD, US}
\author{Matthew Begley}
\address{Department of Mechanical Engineering University of California, Santa Barbara, Santa Barbara, CA, US}
\address{Materials Department, University of California, Santa Barbara, Santa Barbara, CA, US}
\author{Nicholas Boechler\thanks{Corresponding author}*}
\ead{nboechler@ucsd.edu}
\address{Department of Mechanical and Aerospace Engineering, University of California, San Diego, La Jolla, CA, US}
\address{Program in Materials Science and Engineering, University of California, San Diego, La Jolla, CA, US}

\begin{abstract}
Bistable mechanical metamaterials have shown promise for mitigating the harmful consequences of impact by converting kinetic energy into stored strain energy, offering an alternative and potentially synergistic approach to conventional methods of attenuating energy transmission. In this work, we numerically study the dynamic response of a one-dimensional bistable metamaterial struck by a high speed impactor (where the impactor velocity is commensurate with the sound speed), using the peak kinetic energy experienced at midpoint of the metamaterial compared to that in an otherwise identical linear system as our performance metric. We make five key findings: 1) The bistable material can counter-intuitively perform better (to nearly $10^3 \times$ better than the linear system) as the viscosity \emph{decreases} (but remains finite), but only when sufficiently fine discretization has been reached (\textit{i.e.} the system approaches sufficiently close to the continuum limit); 2) This discretization threshold is sharp, and depends on the viscosity present; 3) The bistable materials can also perform significantly worse than linear systems (for low discretization and viscosity or zero viscosity); 4) The dependence on discretization stems from the partition of energy into trains of solitary waves that have pulse lengths proportional to the unit cell size, where, with intersite viscosity, the solitary wave trains induce high velocity gradients and thus enhanced damping compared to linear, and low-unit-cell-number bistable, materials; and 5) When sufficiently fine discretization has been reached at low viscosities, the bistable system outperforms the linear one for a wide range of impactor conditions. The first point is particularly important, as it shows the existence of a nonlinear dynamical ``size effect'', where, given a protective layer of some thickness and otherwise identical quasi-static mechanical properties and total mass, \textit{e.g.}, a $1$ mm thick layer having $200$ unit cells of $5$ $\mu$m thickness is predicted to perform significantly better than one having $20$ unit cells of $50$ $\mu$m thickness. The complex dynamics revealed herein could help guide the future design and application of bistable, and perhaps more generally nonlinear, metamaterials in various domains including signal processing, shape changing devices, and shock and impact protection, with particular benefits in the latter case predicted for scenarios where constituent materials with low intrinsic viscosity are needed (\textit{e.g.}, wherein metals or ceramics would be used). 
\end{abstract}


\end{frontmatter}

\section{Introduction}\label{introduction}

\noindent Reducing the transmission of energy from impacts is crucial in a myriad of natural and engineered systems \cite{meyers1994dynamic,national2011opportunities,sun2009energy,san2020review,lazarus2020review,sun2018dynamic,qiao2008impact}. Viscosity is frequently used for this \cite{lakes2009viscoelastic}, however one limitation is that there is a known trade-off between stiffness (also correlated with strength and toughness) and damping in conventional materials \cite{AshbyM.F1999Msim}. Mechanisms such as plastic deformation \cite{national2011opportunities,sun2009energy,san2020review,lazarus2020review,sun2018dynamic,qiao2008impact} are also used, with the key downside that they are irreversible and thus single use. In ``wave dominated'' regimes (where dominant wavelengths of the impact pulse are smaller than the material size), scattering can also reduce transmitted energy amplitudes \cite{ponson2010nonlinear}, including via interplay with damping \cite{tournat2004multiple}. A newer, alternative approach is to trap energy elastically using bistability \cite{cedolin1991stability} (\textit{e.g.} via beams that snap under compression from one state to another), sometimes within the form of a ``metamaterial'' where repeating arrays of bistable materials are arranged to form an effective, composite-like, material \cite{correa2015negative,shan2015multistable, cao2021bistable}. Bistable elastic energy trapping is reversible, does not abide by the same tradeoffs as viscosity, and can foreeseably be applied synergistically with the aforementioned mechanisms \cite{jeon2022synergistic}. Such bistable mechanisms often coexist with energy absorption via the interplay of negative stiffness, hysteresis, and damping \cite{lakes2001extreme,ding2020designs,harne2013review}. 

Experimental studies on bistable metamaterials demonstrating energy trapping (\textit{e.g.}, Ref. \cite{shan2015multistable}) have focused on low rates where the impactor velocities are much smaller than the material sound speed. In parallel, several computational studies have explored the energy absorbing capacity of bistable mechanical metamaterials and lattices at higher rates and wave dominated regimes \cite{SLEPYAN2005407,cohen2014dynamics,katz2018solitary,katz2019solitary,fancher2023dependence}. In the thesis \cite{fancher2022physics} by the second author of this work, an intriguing, further phenomenon was observed---that the degree of discretization in a damped bistable mechanical metamaterial drastically affected the energy transmission performance, particularly in the presence of damping. This phenomenon was not observed to occur in the comparative, simulated linear system. While the dynamics of damped, discrete nonlinear systems have been well addressed \cite{purwins2010dissipative,chong2017nonlinear,raney2016stable}, this interdependence of kinetic energy transmission on discretization has not been described. Indeed, this phenomenon may be somewhat surprising if imagining increased discretization serving as a better and better approximation of a continuum. 

\begin{figure}[h!]
\centering
  \includegraphics[width=6.9in]{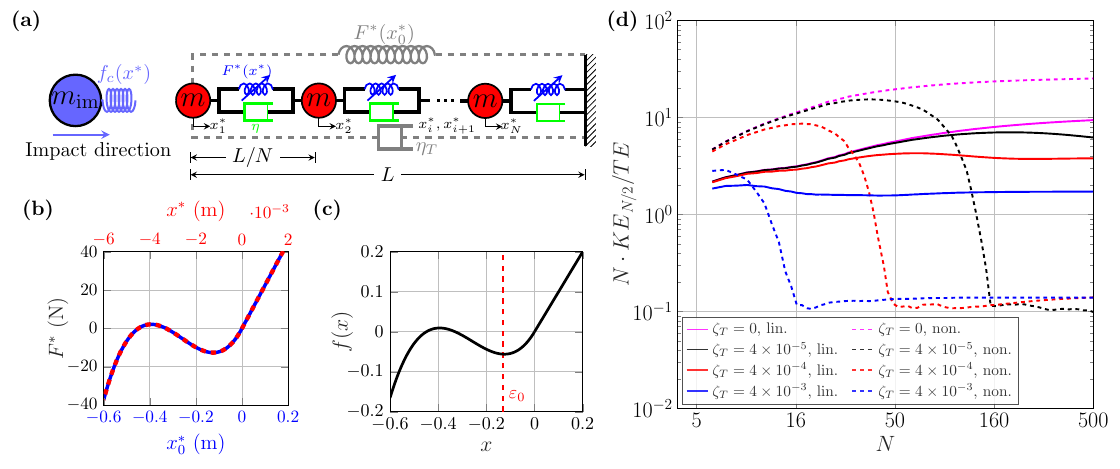}%
\caption{\textbf{Bistable metamaterial model and the first three key findings.} (a) DEM model (b) Bistable force-displacement response of a single layer (red dashed line, top axis) and the entire $N$-layered metamaterial (blue line, bottom axis). (c) Non-dimensionalized single layer force-displacement response. (d) The first three key findings demonstrated in terms of peak kinetic energy experienced at the midpoint of the material ($KE_{N/2}$) normalized by the initial total energy ($TE$) of the system for the linear (solid) and bistable (dashed) materials versus $N$ (unit cells within the system length) for select bulk damping ratios ($\zeta_T$) with impactor mass $M/M_0=2$ and velocity $V/V_0=2$. The first three key findings are: 1) The bistable material can counter-intuitively perform better as the viscosity \emph{decreases} (but remains finite), but only when sufficiently fine discretization has been reached (\textit{i.e.} the system approaches sufficiently close to the continuum limit); 2) This discretization threshold is sharp, and depends on the viscosity present; and 3) The bistable materials can also perform significantly worse than linear systems (for low discretization and viscosity or zero viscosity).}
\label{schematic_bistable_metamateirl}
\end{figure}

For a bistable mechanical metamaterial experiencing high rate impact within wave dominated regimes, we show, via computational analysis an intricate relationship between the transmission of kinetic energy and the discreteness and viscosity of the material (all other bulk properties remaining constant). Using the peak kinetic energy experienced at midpoint of of a one-dimensional (1D) metamaterial compared to that in an otherwise identical linear system as our performance metric, we make five key findings: 1) The bistable material can counter-intuitively perform better (to nearly $10^3 \times$ better than the linear system) as the viscosity \emph{decreases} (but remains finite), but only when sufficiently fine discretization has been reached (\textit{i.e.} the system approaches sufficiently close to the continuum limit); 2) This discretization threshold is sharp, and depends on the viscosity present; 3) The bistable materials can also perform significantly worse than linear systems (for low discretization and viscosity or zero viscosity); 4) The dependence on discretization stems from the partition of energy into trains of solitary waves (localized pulses that maintain their shape due to a balance of nonlinearity and dispersion \cite{dauxois2006physics}) with pulse widths proportional to the unit cell size, where, with intersite viscosity, this feature of decomposition of an impact pulse into solitary wave trains yields high velocity gradients and thus enhanced damping compared to linear, and low unit-cell number bistable, materials; and 5) When sufficiently fine discretization has been reached at low viscosities, the bistable system outperforms the linear one for a wide range of impactor conditions. The first point is particularly important, as it shows the existence of a nonlinear dynamical ``size effect'', where, given a protective layer of some thickness and otherwise identical quasi-static mechanical properties and total mass, \textit{e.g.}, a $1$ mm thick layer having $200$ unit cells of of $5$ $\mu$m thickness is predicted to perform significantly better than one having $20$ unit cells of $50$ $\mu$m thickness. The computational model for the mechanical metamaterial is a discrete element model (DEM), featuring a 1D chain of masses interconnected by bistable nonlinear springs and intersite damping, shown in Fig. \ref{schematic_bistable_metamateirl}(a-c). The first two key findings can be seen in Fig. \ref{schematic_bistable_metamateirl}(d), which shows the kinetic energy transmitted to the midpoint of the material for a given impact as a function of damping ratio ($\zeta_T$) and discretization, where $N$ is the number of unit cells within the system length (such that $N\to\infty$ is the continuum limit). We suggest that the complex dynamical interplay between viscosity, nonlinearity, and discreteness revealed herein could help guide the future design and application of bistable and, more generally, nonlinear, metamaterials. This includes potential uses not just in shock and impact protection, but also more broadly as signal processing and shape changing devices. As regards impact protection, we note that the viscosity range considered herein corresponds to constituent materials such as higher damping ceramics and low damping metals. 

\section{Model}\label{theory}

\noindent The DEM of the energy absorbing material considered is that of a rigid ``impactor'' striking a lumped mass lattice material with total length of $L$ and intersite bistable springs (Fig. \ref{schematic_bistable_metamateirl}(a)). The force, $F^\ast$, and displacement, $x^\ast$, relation of the bistable spring used is from Ref. \cite{fancher2023dependence} (wherein it was fit from the measured response of a 3D printed unit cell of bistable metamaterial), and takes the form 
\begin{equation}\label{force_disp_unitcell}
F^\ast = c_{r1}^\ast x^\ast+c_{r2}^\ast [x^\ast]_+^2+c_{r3}^\ast [x^\ast]_+^3, \quad x^\ast\in[-a,\infty],
\end{equation}
where $F^\ast>0$ and $x^\ast>0$ denotes tension and extension, respectively, the operator $[\cdot]_+$ denotes zero in tension, $a$ is the unit cell length, and the coefficients $c_{r1}^\ast=2.257\times10^4$ N/m, $c_{r2}^\ast=1.187\times10^7$ N/m, and $c_{r3}^\ast=1.524\times10^9$ N/m (plotted in Fig. \ref{schematic_bistable_metamateirl}(b)). The superscript $^\ast$ is used to represent dimensional quantities for which we later define dimensionless counterparts. In Ref. \cite{fancher2023dependence}, the entire simulated impact absorbing metamaterial is 100 layers, such that the effective nonlinear force-displacement response of the entire bulk (with $x_0^\ast$ the total quasi-static displacement of the bulk) is thus 
\begin{equation}\label{force_disp_bulk}
F^\ast = c_{01}^\ast x_0^\ast+c_{02}^\ast [x_0^\ast]_+^2+c_{03}^\ast [x_0^\ast]_+^3 \quad x_0^\ast\in[-L,\infty] \\
\end{equation}
where $c_{01}^\ast=c_{r1}^\ast/100=225.7$ N/m, $c_{02}^\ast=c_{r2}^\ast/100^2=1187$ N/m, $c_{03}^\ast=c_{r3}^\ast/100^3=1524$ N/m. We further denote the linear stiffness of the bulk as $k_T=c_{01}^\ast$. For the remainder of the study, we keep these bulk mechanical properties constant ($L$, $k_T$, and total mass $m_T$). In Appendix A, we validate that the bulk properties are maintained for quasi-static deformation in our computational model.

We next assume the bistable metamaterial has bulk damping $\eta_T$ and consists of $N$ unit cells. The force, displacement, and velocity relation of a unit cell can thus be written by the bulk parameters and the number of unit cells, such that
\begin{equation}\label{force_disp_Nunit}
\begin{split}
F^\ast (x^\ast, \dot{x^\ast})&=c_1^{\ast}x^{\ast}+c_2^{\ast}[x^{\ast}]_+^2+c_3^{\ast}[x^{\ast}]_+^3+\eta\dot{x^\ast}\quad x^\ast\in[-a,\infty]\\
&=c_{01}^{\ast}x_0^{\ast}+c_{02}^{\ast}[x_0^{\ast}]_+^2+c_{03}^{\ast}[x_0^{\ast}]_+^2+\eta_T\dot{x_0^\ast}\quad x_0^\ast\in[-L,\infty],
\end{split}
\end{equation}
where $x^{\ast}=x_{0}^{\ast}/N$, $c_1^{\ast}=Nc_{01}^{\ast}$,  $c_2^{\ast}=N^2c_{02}^{\ast}$, $c_3^{\ast}=N^3c_{03}^{\ast}$, and $\eta = N \eta_T$. 

\subsection{DEM and non-dimensionalization}\label{dem_non}

\noindent Using the bistable spring response of Eq.~\ref{force_disp_Nunit}, the equation of motion for $i$th unit cell can be written as 
\begin{equation}\label{dem_dimension}
\begin{split}
&m\ddot{x}^\ast_{i}-c_1^\ast(x^\ast_{i+1}-x^\ast_{i})-c_2^\ast[x^\ast_{i+1}-x^\ast_{i}]_+^2-c_3^\ast[x^\ast_{i+1}-x^\ast_{i}]_+^3\\
&+c^\ast_1(x^\ast_{i}-x^\ast_{i-1})+c_2^\ast[x^\ast_{i}-x^\ast_{i-1}]_+^2+c_3^\ast[x^\ast_{i}-x^\ast_{i-1})]_+^3+\eta(-\dot{x}^\ast_{i+1}+2\dot{x}^\ast_{i}-\dot{x}^\ast_{i-1})=0,
\end{split}
\end{equation}
where $m=m_T/N$ is the mass of the unit cell and $i\in[1,2,...,N]$. We then normalize by bulk parameters as follows: 
\begin{equation}\label{nondimensional_variables}
\begin{split}
  &  x=\frac{Nx^\ast}{L},\quad t=\frac{Nt^\ast}{\sqrt{m_T/k_T}},\quad \frac{dx}{dt}=\frac{\sqrt{m_T/k_T}}{L}\frac{dx^\ast}{dt^\ast}, \quad \frac{d^2x}{dt^2}=\frac{m_T}{Nk_TL}\frac{d^2x^\ast}{dt^{\ast2}}\\
  &\zeta_T=\frac{\eta_T}{2\sqrt{m_Tk_T}},\quad \zeta=N\zeta_T, \quad c_2=\frac{c_2^\ast L}{N^2k_T}=\frac{c_{02}^\ast L}{k_T},  \quad c_3=\frac{c_3^\ast L^2}{N^3k_T}=\frac{c_{03}^\ast L^2}{k_T}, \quad  f(x)=\frac{F(x^\ast)}{k_TL}.
\end{split}
\end{equation}
The nondimensional equation of motion of the $i$-th unit cell using bulk parameters is thus
\begin{equation}\label{dem_nondimension}
\begin{split}
   & \ddot{x}_{i}-(x_{i+1}-x_{i})-c_2[x_{i+1}-x_{i}]_+^2-c_3[x_{i+1}-x_{i}]_+^3\\
   &+(x_{i}-x_{i-1})+c_2[x_{i}-x_{i-1}]_+^2+c_3[x_{i}-x_{i-1}]_+^3+2\zeta(-\dot{x}_{i+1}+2\dot{x}_{i}-\dot{x}_{i-1})=0,
\end{split}
\end{equation}
where $f(x)=x+c_2[x]_+^2+c_3[x]_+^3$ is the dimensionless nonlinear spring force with $c_2=5.2592$, $c_3=6.7523$, and $x=(x_{i+1}-x_{i})\in[-1,\infty]$, as is shown in Fig. \ref{schematic_bistable_metamateirl}(c). 

A Hertzian contact \cite{hertz1882ueber} spring between the left unit cell and the impactor mass is modeled to describe the impactor hitting the the sample and freely bouncing, rather than sticking, as well as smooth onset and release of contacts, such that 
\begin{equation}\label{contact_force}
 f_c^\ast(x^\ast_{im},x^\ast_1)=C_{imp}\sqrt{[x^\ast_{im}-x^\ast_{1}]_+^3},
\end{equation}
where $C_{imp}=2\times 10^9$ is chosen. It is important to note that given this definition, the contact spring does not change with discretization $N$. The nondimensional contact force can further be written as
\begin{equation}\label{nondimension_contact_force}
  f_c(x_{im},x_{1})=\frac{C_{imp}\sqrt{L}}{k_T\sqrt{N^3}}\sqrt{[x_{im}-x_{1}]_+^3}.
\end{equation}

The DEM thus consists of a system of $N+1$ equations of motion written as:
\begin{equation}\label{dem_impact}
\begin{split}
\begin{cases}
&NM\ddot{x}_{im}+f_c(x_{im},x_{1})=0\\
&\ddot{x}_{i}-(x_{i+1}-x_{i})-c_2[x_{i+1}-x_{i}]_+^2-c_3[x_{i+1}-x_{i}]_+^3+2\zeta(-\dot{x}_{i+1}+\dot{x}_{i})-f_c(x_{im},x_{i})=0 \quad(i=1)\\
&\ddot{x}_{i}-(x_{i+1}-x_{i})-c_2[x_{i+1}-x_{i}]_+^2-c_3[x_{i+1}-x_{i}]_+^3+(x_{i}-x_{i-1})+c_2[x_{i}-x_{i-1}]_+^2+c_3[x_{i}-x_{i-1}]_+^3\\
&+2\zeta(-\dot{x}_{i+1}+2\dot{x}_{i}-\dot{x}_{i-1})=0 \quad(2\leqslant i \leqslant N-1)\\
&\ddot{x}_{i}+x_{i}+c_2[x_{i}]_+^2+c_3[x_{i}]_+^3+(x_{i}-x_{i-1})+c_2[x_{i}-x_{i-1}]_+^2+c_3[x_{i}-x_{i-1}]_+^3+2\zeta(2\dot{x}_{i}-\dot{x}_{i-1})=0  \quad(i=N),
\end{cases}
\end{split}
\end{equation}
where $M=m_{im}/m_T$.

\subsection{Nominal impact conditions}\label{nominal_impact}

\noindent As in Ref.~\cite{fancher2023dependence}, we estimate the nominal impact velocity as
\begin{equation}\label{nominal_velocity}
  V_0^\ast=2\varepsilon_0c_0,
\end{equation}
where $c_0=L\sqrt{k_T/m_T}$ is the long wavelength linear sound speed and $\varepsilon_0=\abs{-0.13}$ is the first inflection point of the strain energy vs. displacement curve  (as per Fig. \ref{schematic_bistable_metamateirl}(c)). We then normalize the impact velocity by $c_0$ such that $V_0=2\varepsilon_0$. Equation (\ref{nominal_velocity}) can be considered to estimate the required velocity to induce snap-through of the material to the second stable state.

Herein, we choose, following Ref.~\cite{fancher2023dependence}, our lattice as our performance figure of merit as the kinetic energy measured at the middle particle. Thus, with the goal of absorbing the impactor's kinetic energy in the first half of the sample we, still following Ref.~\cite{fancher2023dependence}, define a nominal impactor mass, $M_0^\ast$, by setting the impactor kinetic energy equal to the elastic strain energy produced by compression of half the unit cells in the bistable metamaterial to the second stable state ($\abs{\varepsilon_2}=0.46$), such that
\begin{equation}\label{nominal_mass_dimensional}
\frac{1}{2}M_0^\ast V_0^{\ast2}=\frac{N}{2}\int_{0}^{\varepsilon_2 a}(c_1^\ast x^{\ast}-c_2^\ast x^{\ast2}+c_3^\ast x^{\ast3})dx^{\ast}.
\end{equation}

\noindent Defining a nondimensional impactor mass $M_0=M_0^\ast/m_T$, we can rewrite Eq.~\ref{nominal_mass_dimensional} in terms of nondimensional quantities as \begin{equation}\label{nominal_mass}
  \frac{1}{2}M_0V_0^2=\frac{1}{2}\int_{0}^{\varepsilon_2}(x-c_2x^2+c_3x^3)dx.
\end{equation} 
The nominal impactor mass $M_0$ can then be solved for as the only remaining unknown in Eq. (\ref{nominal_mass}) to be $M_0=0.159$ (giving $M_0^\ast=0.159m_T$).

Given these nominal impactor conditions, it may be useful to comment on the practical setting which we are considering herein. The nominal impactor velocity would be $26 \%$ of the sound speed of the bistable mechanical metamaterial and the impactor would be $15.9 \%$ of the total system's mass. Considering first the impactor velocity, we emphasize this is relative to the effective sound speed of the bistable material, not that of the constituent materials from which it is made. Due to the structural inefficiency of adding porosity  \cite{zheng2014ultralight} (\textit{e.g.} in the form of bistable structures), this results in the effective sound speed being on the order, but less than that of the constituent material. The effective sound speed could indeed be drastically lower than that of the constituent material, but that would imply there is reduced capacity to store strain energy in bistable mechanisms. As such, considering, for example, a metal constituent material (sound speed $\sim 5000$ m/s), one could imagine impactor velocities $\sim 1200$ m/s, whereas for in a soft elastomer block, wavespeeds could be as low as $\sim 44$ m/s \cite{abi2019wrinkles}, giving impactor velocities of $\sim 11$ m/s. For a constituent material such as metals experiencing such high impact speed, plastic mechanisms and constituent material strain rate dependency could be expected, however we do not consider these mechanisms herein. Considering next the mass, since the nominal impactor mass scales with the total system mass, one can imagine lighter nominal impactor masses will be more appropriate for thinner protective layers or lower density (typically softer) constituent materials undergoing impact.

\section{Results}\label{numerical}

\begin{figure}[h!]
\centering
  \includegraphics[width=6.8in]{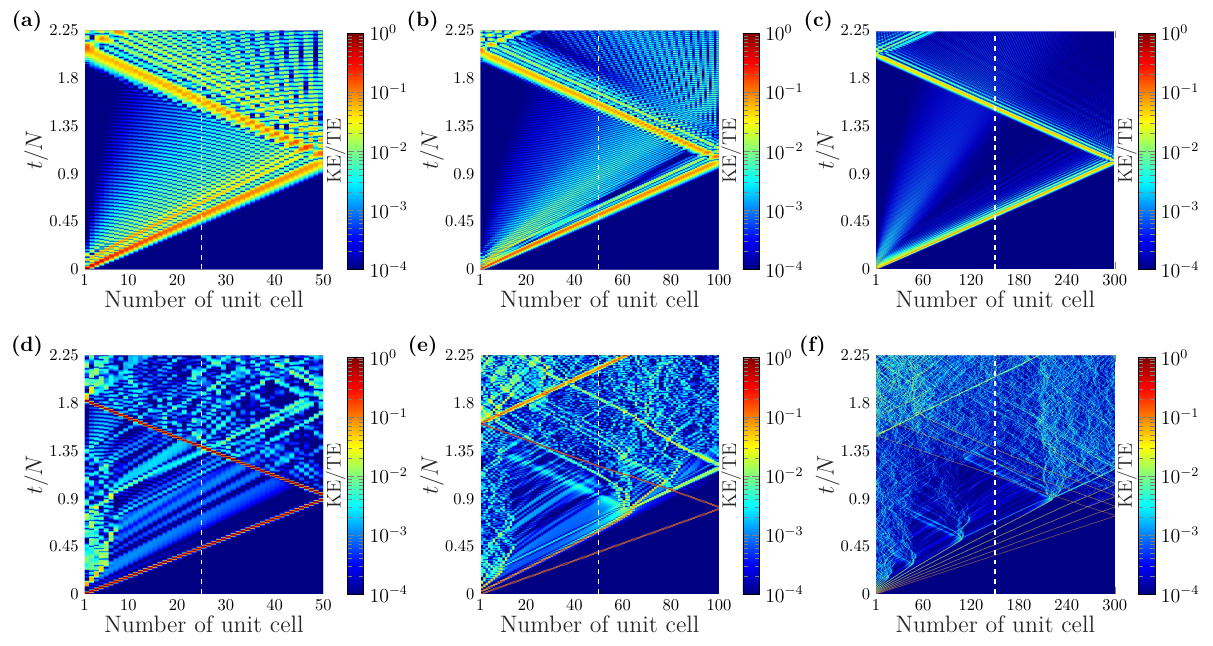}%
\caption{\textbf{Effect of changing discretization on the response of undamped linear (top row, a-c) and bistable (bottom row, d-f) systems}. Kinetic energy for each particle normalized by the initial total energy of the system (initial impactor kinetic energy), denoted as KE/TE, for impact conditions of $M/M_0=0.5$ and $V/V_0=4$. (a,d) $N=50$, (b,e) $N=100$, and (c,f) $N=300$. The physical time is identical for all $N$ cases. Suboptimal impact conditions were chosen to highlight the qualitative difference between the linear and bistable material responses.}\label{KE_m05v4_N_undamped}
\end{figure}

\noindent We numerically integrate Eq. (\ref{dem_impact}) using a Runge-Kutta algorithm (MATLAB's \textit{ode45()})with initial nondimensional velocity $V$ applied to the impactor particle (where $V=V^\ast/c_0$, with $V^\ast$ the dimensional impactor velocity). In Fig. \ref{KE_m05v4_N_undamped}, we show the effect of varying the discretization $N$ with a given impact condition ($M/M_0=0.5$ and $V/V_0=4$) for a undamped linear (top row) and bistable mechanical metamaterial (bottom row). We note that all heatmap plots in this manuscript are plotted using the \textit{pcolor()} function in MATLAB. All spatiotemporal response plots in this manuscript plot kinetic energy for each particle normalized by the initial total energy of the system, wherein the color bars are set to a minimum of $10^{-4}$. In the linear material, all parameters are the same as the simulated bistable mechanical metamaterial except the bistable spring is replaced with a linear one set to match the linear stiffness term of the bistable spring. In Fig. \ref{KE_m05v4_N_undamped}, a vertical white dashed line denotes the middle particle in our system, wherein we aim to transmit the least kinetic energy possible. In Fig. \ref{KE_m05v4_N_undamped}, for the linear system (top row), we see a clear pulse propagate across the material and reflect from the opposite side, where the primary effect that can be seen with increasing discretization is a reduction in dispersion of the pulse. In contrast, in the bistable material (bottom row of Fig. \ref{KE_m05v4_N_undamped}), we see a clear qualitative change in the dynamic response, where, as $N$ increases, the impact is broken up into ``trains'' of more localized solitary waves. This solitary wave train phenomena has been previously studied in other nonlinear discrete systems such as granular chains \cite{nesterenko2013dynamics,job2007solitary,sokolow2007solitary}. Indeed, two features consistent with the dynamic response of granular chains can be observed in the bottom row of Fig.~\ref{KE_m05v4_N_undamped}. First, in granular chains, when the mass of the impactor is greater than the mass of an individual particle, the response transitions from a single solitary wave to a train of solitary waves where each subsequent wave in the train has decreasing energy compared to the first \cite{nesterenko2013dynamics,job2007solitary,sokolow2007solitary}. Here, as $N$ increases, the mass of the individual particles constituting the chain drops with respect to that of the impactor. However, in contrast to granular chains, based on the bottom row of Fig.~\ref{KE_m05v4_N_undamped}, the transition from single to multiple solitary waves appears to be between $3.975<m_{im}/m<7.95$, corresponding to panels (d) and (e). We note that we cannot conclude that impactor velocity does not play a role in the transition from these initial cases, as would be the case for granular chains. The second similar feature to granular chains and other nonlinear systems that support solitary waves \cite{dauxois2006physics} is that the solitary waves in the bistable system exhibit speed proportional to their amplitude (\textit{i.e.} the first, highest amplitude wave in the train has the greatest speed, with each subsequent in the train traveling slower and slower. Returning to the comparison between the overall linear and bistable system response, we note that the performance (peak kinetic energy experienced at the midpoint) of the bistable system shown in Fig. \ref{KE_m05v4_N_undamped} is worse than that of the linear system. These suboptimal impact conditions were chosen to highlight qualitative change from the linear pulses to the nonlinear solitary wave trains. When comparing the performance at different discretizations, it is important to note that the kinetic energy in the scale bar in Fig. \ref{KE_m05v4_N_undamped} is that of the particle, so it is expected that higher discretization (increasing from left ot right columns) results in less kinetic energy per particle. As can be seen in Fig.~\ref{schematic_bistable_metamateirl}(d), when comparing systems with different discretization, this is address by multiplying by $N$, so as to obtain an effective peak kinetic energy per unit length. 

\begin{figure}[h!]
\centering
  \includegraphics[width=6.8in]{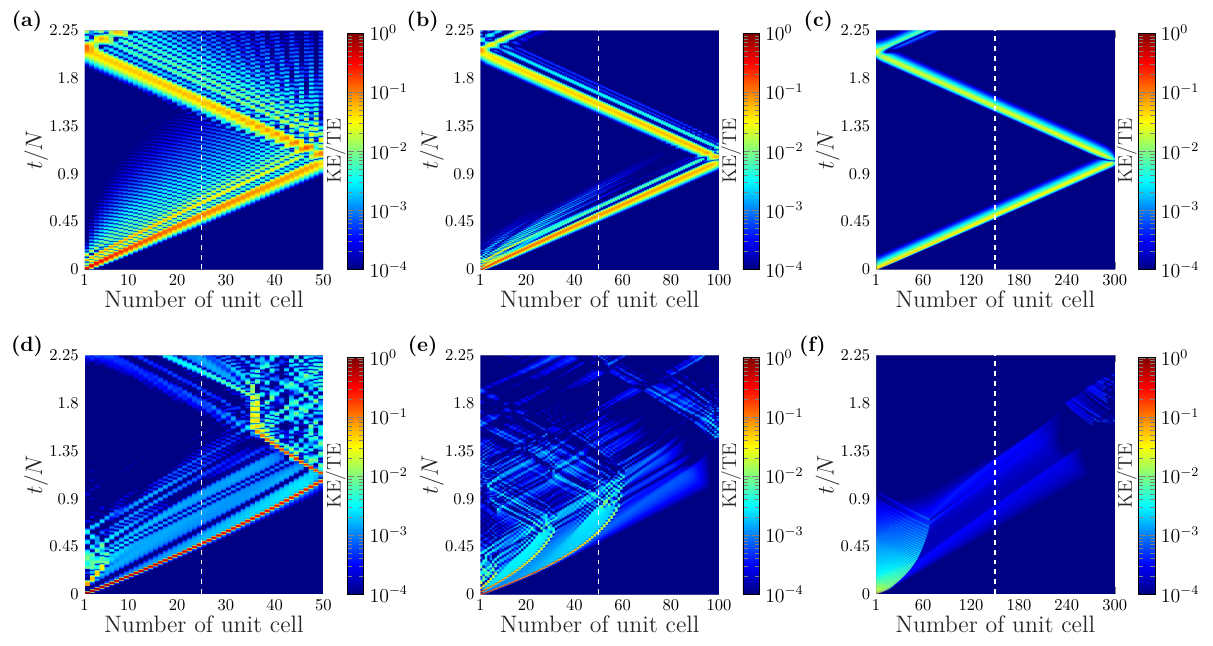}%
\caption{\textbf{Effect of changing discretization on the response of linear (top row, a-c) and bistable (bottom row, d-f) systems for the same impact conditions as Fig.~\ref{KE_m05v4_N_undamped}, but now with damping added}. Kinetic energy for each particle normalized by the initial total energy of the system (initial impactor kinetic energy) for the simulated damped ($\zeta_T=10^{-4}$) linear (a-c) and bistable metamaterial (d-f). (a,d) $N=50$, (b,e) $N=100$, and (c,f) $N=300$. The physical time is identical for all $N$ cases}\label{KE_m05v4_N_damped}
\end{figure}

In Fig.~\ref{KE_m05v4_N_damped}, we show the response for the same system and conditions as in Fig. \ref{KE_m05v4_N_undamped}, but add a bulk viscous damping of $\zeta_T=10^{-4}$. Between Fig.~\ref{KE_m05v4_N_undamped} (undamped) and Fig.~\ref{KE_m05v4_N_damped} (damped), there appears little difference in the linear system response. In the nonlinear system however, there is a clear qualitative difference with the addition of damping. In Fig.~\ref{KE_m05v4_N_damped}, we see that the solitary waves now slow as they propagate (as can be seen by the curved pulse trajectory). This is again consistent with the solitary wave phenomena in granular chains and other nonlinear discrete systems \cite{dauxois2006physics} where, as the amplitude reduces (here, due to the addition of viscous damping), the wave slows. In addition, when the velocity slows sufficiently (\textit{e.g.}, more vertical trajectories in the spatiotemporal plots of Fig.~\ref{KE_m05v4_N_damped}(e)), the pulse disappears and some reflection is induced. One can imagine this is due to insufficient remaining energy in the particular pulse available to open the last bistable spring after it snapped shut, with the leftover energy being reflected backwards. In the highest discretization case (Fig.~\ref{KE_m05v4_N_damped}(f)), the multiple solitary waves within the wave train are so finely spaced that they appear to converge together, and repeated overlapping reflections from earlier pulses can be seen with the later pulses in the train. This results in very little energy being transmitted past the midpoint.

\begin{figure}[h!]
\centering
  \includegraphics[width=6.8in]{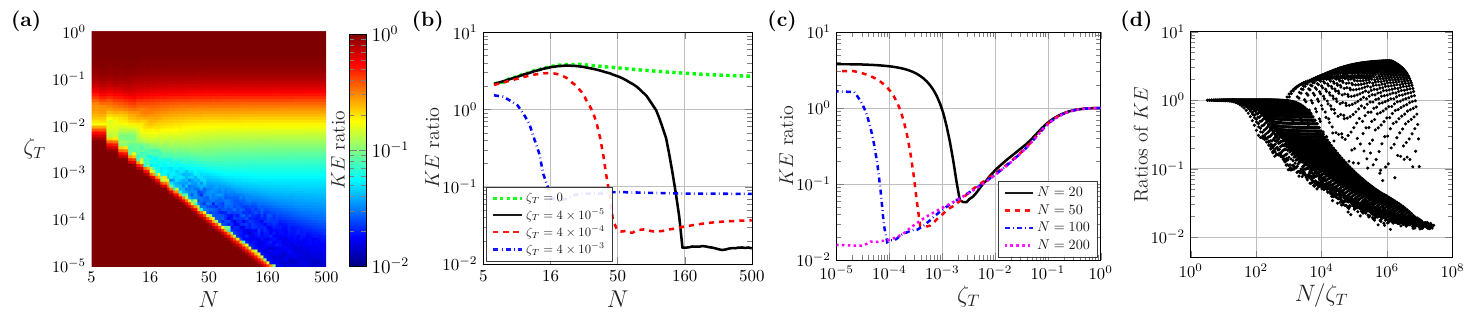}%
\caption{\textbf{Maximum kinetic energy transmitted past the midpoint of the simulated bistable material divided by the maximum kinetic energy transmitted past the midpoint in the comparative linear material (KE ratio) for a fixed impact condition of $\bm{M/M_0=2}$ and $\bm{V/V_0=2}$}. (a) KE ratio vs. both parameters. Slices of (a) at select $\zeta_T$ (b) and $N$ (c). Panel (b) has an added curve for $\zeta_T=0$. (d) KE ratio from (a) vs. $N/\zeta_T$. }\label{KEratio_mv_zetaN_sweeping}
\end{figure}

We next study the dependence of maximum kinetic energy observed at the middle particle of the bistable mechanical metamaterial divided by the maximum kinetic energy at the middle particle of the comparative linear material (again, all parameters are the same as the simulated bistable mechanical metamaterial, except the bistable spring is replaced with a linear one matching the linear stiffness term of the bistable spring), over a time range $t/N=2.25$, on the system and impactor parameters. We refer to this ratio as the ``KE ratio'', wherein the minimum KE ratio would be the best for impact protection. We note that the constant $t/N$ range denotes a fixed physical time, the same of which is used for Figs.~\ref{KE_m05v4_N_undamped} - \ref{KE_m05v4_N_damped}. We first select a single impact condition $M/M_0=2$ and $V/V_0=2$ and vary the bulk damping from $\zeta_T=10^{-5}$ to 1 and the discretization from $N=5$ to $500$, as shown in Fig.~\ref{KEratio_mv_zetaN_sweeping}(a). The KE ratio is truncated in Fig.~\ref{KEratio_mv_zetaN_sweeping}(a) on the high side at unity, and the color bar is set to a minimum of $10^{-2}$, noting that the minimum KE ratio found for this range of parameters is $0.0130$ at $\zeta_T=3.3\times10^{-5}$ and $N=191$. In Fig.~\ref{KEratio_mv_zetaN_sweeping}(a), we see that the best (relative) performance occurs at high $N$ and low damping ($\zeta_T$). This is consistent with the behavior shown in Fig.~\ref{schematic_bistable_metamateirl}(d). Considering $\zeta_T$, Fig.~\ref{KEratio_mv_zetaN_sweeping}(a) shows that this nonlinear elastic, bistable mechanism is advantageous compared to the comparative linear system in cases where the damping is sufficiently low. However, there is a caveat to this, namely that considering both parameters together, a threshold of sufficiently fine discretization (high $N$) is needed to see the performance advantage at low $\zeta_T$ (again as also shown in Fig.~\ref{schematic_bistable_metamateirl}(d)). This threshold can be seen by the diagonal line going from the bottom middle-right to middle left of Fig.~\ref{KEratio_mv_zetaN_sweeping}(a) (noting that Fig.~\ref{KEratio_mv_zetaN_sweeping}(a) is plotted in log scale on both axes). Given that the threshold forms a diagonal line, we see that after a sufficiently high level of damping is reached ($\sim \zeta_T=10^{-2}$), the transistion would no longer be evident, as viscous dissipation will dominate the system response. At the other extreme, we expect the threshold of discretization $N$ would continue to increase with decreasing damping, noting that our parameters extend to the damping region expected for lower damping metals and higher damping ceramics \cite{lakes2009viscoelastic}.  

We further elucidate this effect by taking slices of Fig.~\ref{KEratio_mv_zetaN_sweeping}(a) at select $\zeta_T$ (Fig.~\ref{KEratio_mv_zetaN_sweeping}(b)) and select $N$ (Fig.~\ref{KEratio_mv_zetaN_sweeping}(c)), observing an over two order of magnitude performance increase from worst to best cases. In Fig.~\ref{KEratio_mv_zetaN_sweeping}(b), we also plot the KE ratio for the undamped bistable material, and see that for these particular impact conditions the undamped bistable material underperforms the linear material for all N (as also shown in Fig.~\ref{schematic_bistable_metamateirl}(d)). 
In Fig.~\ref{KEratio_mv_zetaN_sweeping}(c), we see that for a given damping value, the KE ratio performance of the lower N value may be similar or equal but will not exceed the KE ratio performance of the higher N value, and for lower damping values the higher N discretization outperforms the lower N values, in some cases substantially. This point can also be seen in Fig.~\ref{KEratio_mv_zetaN_sweeping}(a) and Fig.~\ref{schematic_bistable_metamateirl}(d). 

Observing that high $N$ and low, but finite, $\zeta_T$ give the lowest KE ratio (best performance) in Fig.~\ref{KEratio_mv_zetaN_sweeping}(a,c), we plot the KE ratio as a function of $N/\zeta_T$ in Fig.~\ref{KEratio_mv_zetaN_sweeping}(d), which highlights how the best KE ratio varies with increasing $N/\zeta_T$. We also observe a threshold (for this impactor condition and range of damping and discretization) at $\sim N/\zeta_T=10^3$ below which the bistable system exclusively performs better than the comparative linear system, but the magnitude of improvement is limited. This corresponds to the high $\zeta_T$ regions in Fig.~\ref{KEratio_mv_zetaN_sweeping}(c). Above this threshold, the system can perform either significantly better (KE ratio $<1$) or worse (KE ratio $>1$) than the comparative linear system. This bimodal behavior can be seen as a slice at constant, small values of $\zeta_T$ in Fig.~\ref{KEratio_mv_zetaN_sweeping}(a) or (c). 

\begin{figure}[h!]
\centering
 \includegraphics[width=6.8in]{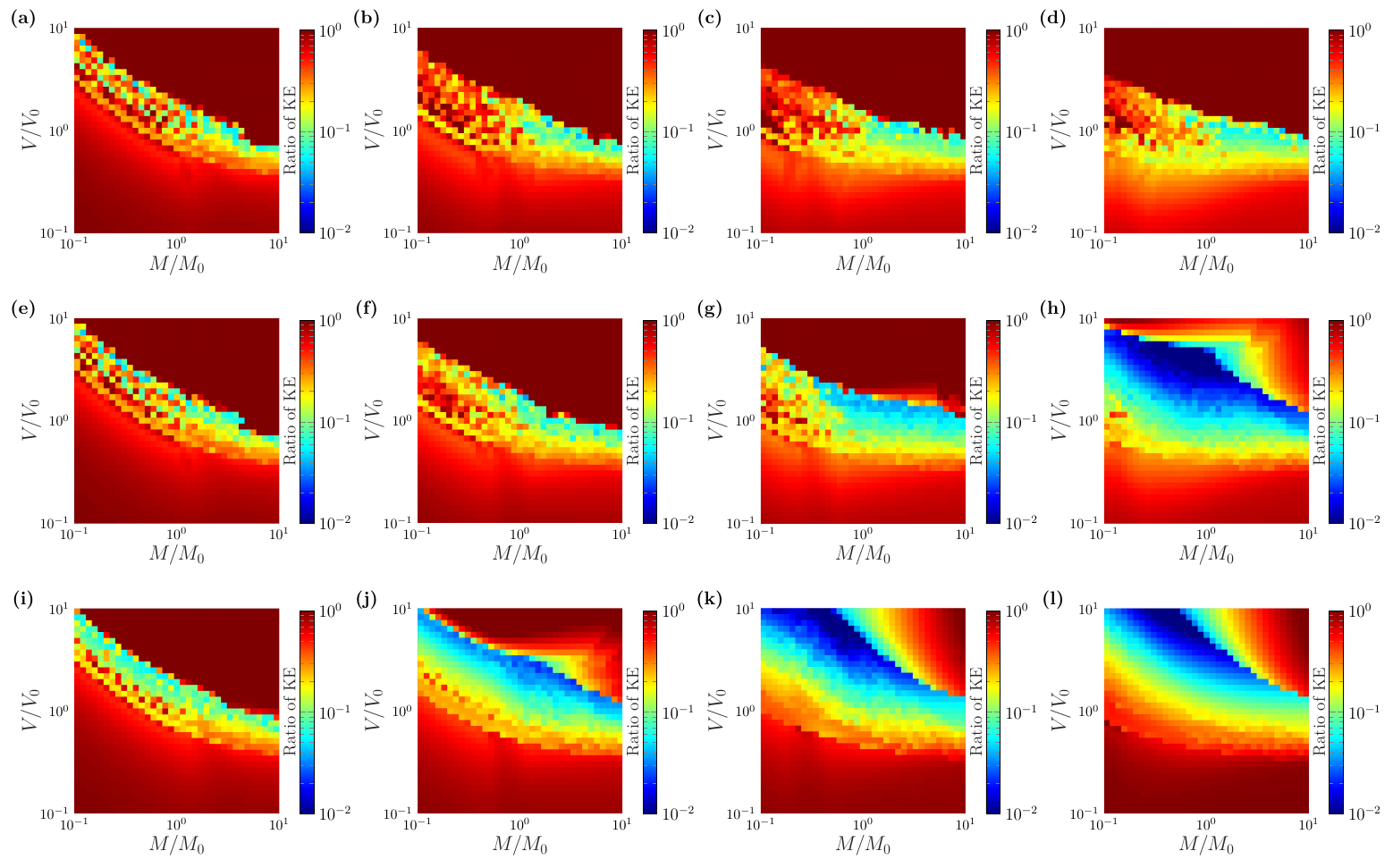}%
\caption{\textbf{KE ratio as a function of impactor mass and velocity for select bulk damping ($\bm{\zeta_T}$) and discretizations ($\bm{N}$)}. [Top row (a-d)] $\zeta_T=0$, [Middle row (e-h)] $\zeta_T=4\times10^{-5}$, and [Bottom row (i-l)] $\zeta_T=4\times10^{-4}$. [Left column (a,e,i)] $N=20$, [Second from left column (b,f,j)] $N=50$, [Second from right column (c,g,k)] $N=100$, [Right column (d,h,l)] $N=200$. The color bars are manually set to a maximum and minimum of $1$ and $10^{-2}$, respectively.}\label{Fig4combo}
\end{figure}

One might ask if the undamped bistable system always underperforms the linear system. In Fig.~\ref{Fig4combo} we choose several damping ratios ($\zeta_T$, constant for a given row) and discretizations ($N$, constant within each column) and plot the KE ratio for a range of impactor conditions ($M/M_0=10^{-1}$ to $10^{1}$ and $V/V_0=10^{-1}$ to $10^{1}$). The top row of Fig.~\ref{Fig4combo} correspond to no damping, and two key features can be seen. First, the bistable system does have a relatively narrow region of impactor conditions where it outperforms the linear system. So the underperformance of the bistable system seen in Fig.~\ref{KEratio_mv_zetaN_sweeping}(b) and Fig.~\ref{schematic_bistable_metamateirl}(d) is not universal for all impact conditions. Second, the minimum of KE ratio (best performance) is fairly limited $\sim 10^{-1}$ to $10^{-2}$ compared to that of the damped systems (once a critical discretization threshold has been reached). 

\begin{figure}[h!]
\centering
 \includegraphics[width=5.5in]{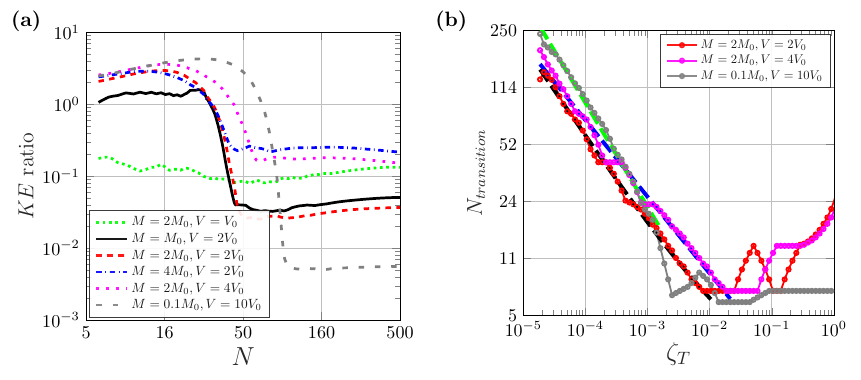}%
\caption{\textbf{Dependence of discreteness transition on impactor mass and velocity, with (a) $\bm{\zeta_T=4\times10^{-4}}$ and (b) varied $\bm{\zeta_T}$}. (a) KE ratio versus discreteness $N$. (b) The critical value $N_{transition}$ versus the bulk damping ratio at which the KE ratio of the materials abruptly shifts from high to low. The black, blue, and green dashed lines are lines of best fit to the magenta, red, and grey datasets, respectively. }\label{KEratio_transition}
\end{figure}

We also ask if changing impactor conditions changes the observed phenomena of: minimized KE transmission compared to the linear system at small damping ($\zeta_T$) but fine discretization (high $N$), the transition to minimized KE transition occurs at higher $N$ as $\zeta_T$ decreases, and that the transition is sharp. 
In Fig. \ref{KEratio_transition}(a), we compare the same quantities as in Fig.~\ref{KEratio_mv_zetaN_sweeping}(b) (and horizontal slices of Fig.~\ref{KEratio_mv_zetaN_sweeping}(a), but for varied impact conditions. We show that, for several impact conditions, the qualitative behavior of KE ratio as a function of discretization is minimally affected by changing impactor mass and velocity, albeit the discretization at which the transition occurs (which we will refer to as $N_{transition}$) appears to shift, as well as the KE ratios before and after the transition. The shifting of the pre- and post-transition KE ratio levels are consistent with the bottom row of Fig.~\ref{Fig4combo}, where it can be seen that, by choosing a specific impact condition and translating between panels (corresponding to discretizations), the KE ratios before and after the transition are different as compared to another chosen impact condition. We further note the KE ratio minima in the high discretization cases for Fig.~\ref{Fig4combo}, namely panel (h) corresponding to $\zeta_T=4\times10^{-5}$ and panel (l) corresponding to $\zeta_T=4\times10^{-4}$, which are $0.0054$ ($M/M_0=0.936$ and $V/V_0=5.179$) and $0.0052$ ($M/M_0=0.553$ and $V/V_0=10$), respectively. In Fig.~\ref{KEratio_transition}(a), we see the effect of changing impactor conditions on $N_{transition}$, where, for the cases chosen, increasing impactor velocity appears to shift $N_{transition}$ higher, with minimal effect from impactor mass. 

\begin{figure}[t!]
\centering
  \includegraphics[width=4.6in]{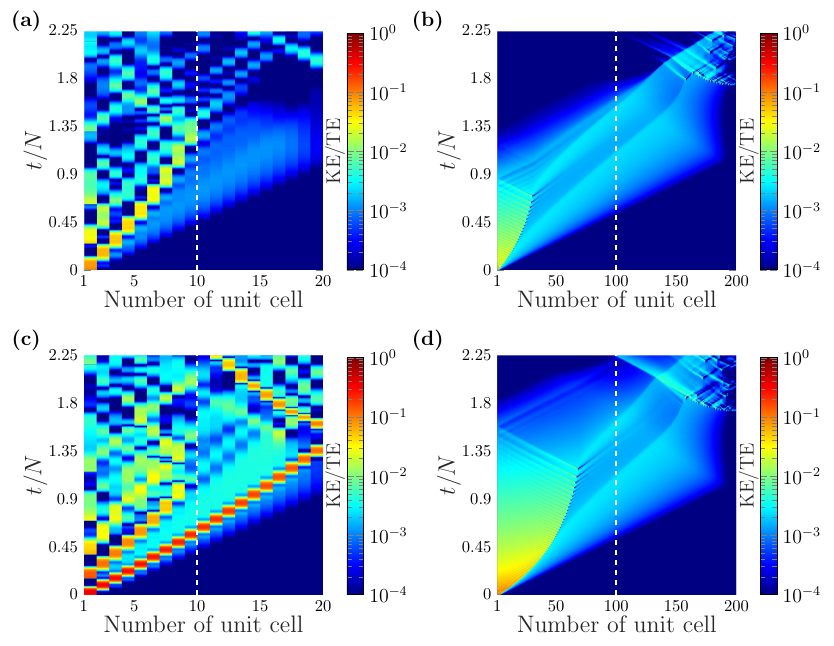}%
\caption{\textbf{Low impactor energy spatiotemporal response}. Kinetic energy for each particle normalized by the initial total energy of the system, denoted as KE/TE, (a,b) $M/M_0=2$, $V/V_0=1$, and (c,d) $M/M_0=2$, $V/V_0=2$. (a,c) $N=20$ and (b,d) $N=200$. The damping ratio are all $\zeta_T=4\times10^{-4}$.}\label{SI_2_KE_m2v1eta_m2v2eta}
\end{figure}

The key exception to the aforementioned similar qualitative response between the varied impactor conditions in Fig.~\ref{KEratio_transition}(a) is the green curve ($M/M_0=2$ and $V/V_0=1$), which corresponds to the lowest impactor energy considered in the figure. In this case, the KE ratio is relatively constant ($\sim 10^{-1}$) with discretization, such that there is not a region where the bistable material performs worse than the linear one. We suggest an explanation via Fig.~\ref{SI_2_KE_m2v1eta_m2v2eta}, where we compare kinetic energy spatiotemporal responses corresponding to the green curve in Fig.~\ref{KEratio_transition}(a) (top row in Fig.~\ref{SI_2_KE_m2v1eta_m2v2eta}) and the red curve in Fig.~\ref{KEratio_transition}(a) (bottom row in Fig.~\ref{SI_2_KE_m2v1eta_m2v2eta}), where the left column denotes a discretization of $N=20$ and the right column $N=200$. The difference for the improved performance and lack of transition can be seen in the $N=20$ cases, where the lower energy case Fig.~\ref{SI_2_KE_m2v1eta_m2v2eta}(a) does not see significant slowing of the solitary waves, but yet the leading pulse stops abruptly and is partially reflected backwards, which corresponds to insufficient energy to snap subsequent cells. We note the peak kinetic energy transmission is at $t/N\approx1.4$, corresponding to the bright blue region to the right of the white dashed line in Fig.~\ref{SI_2_KE_m2v1eta_m2v2eta}(a). In the higher energy case at the same discretization Fig.~\ref{SI_2_KE_m2v1eta_m2v2eta}(b), a solitary wave propagates across the entire material. In the $N=200$ cases (Fig.~\ref{SI_2_KE_m2v1eta_m2v2eta} right column), we see in both cases significant slowing of the solitary waves, suggesting a corresponding significant reduction in pulse amplitude with propagation. As such, for the low energy, low discretization case of Fig.~\ref{SI_2_KE_m2v1eta_m2v2eta}(a), the influence of damping appears to play a minimal role, such that elastic energy trapping is the major deterrent of kinetic energy propagation, until a sufficiently large energy input (\textit{e.g.}, Fig.~\ref{SI_2_KE_m2v1eta_m2v2eta}(b)). We note these regions also correspond to the bottom left front of the high performance band in Fig.~\ref{Fig4combo}.

At the opposite side of the phase space, at sufficiently high kinetic energies, the system also transitions to a regime where the impactor has so much energy the elastic and viscous forces are negligible and the system is effectively crushed or strong solitary waves are generated. This has also previously been shown in Ref.~\cite{fancher2023dependence}, and can be seen in the top right region of all panels in Fig.~\ref{Fig4combo}. One additional feature that can be observed in Fig.~\ref{Fig4combo} is that the range of impactor conditions over which the bistable material outperforms the linear one grows when sufficiently fine discretization is reached (\textit{e.g.}, panels (h), (k), and (l).  

In Fig. \ref{KEratio_transition}(b), we isolate $N_{transition}$ as a function of damping for several impact conditions. We see that in there is a approximately a linear relationship in the log-log scale of Fig.~\ref{KEratio_transition}(b) between $\zeta_T$ and $N$, which equates to a power law relationship with exponent $0.3627$, $0.4574$ and $0.6212$ corresponding to the $M/M_0=2$ and $V/V_0=2$ (red data set, black dashed fitting line), $M/M_0=2$ and $V/V_0=4$ (magenta data set, blue dashed fitting line) and $M/M_0=0.1$ and $V/V_0=10$ (gray data set, green dashed fitting line), respectively. The $N_{transition}$ is identified by taking the derivative of KE ratio vs. $N$ curves, as per Fig. \ref{KEratio_transition}(a), for varied $\zeta_T$. We then manually choose the discretization with the highest amplitude derivative in the parameter region of the transition (\textit{e.g.}, as per the diagonal line in Fig.~\ref{KEratio_mv_zetaN_sweeping}(a)). The observed power law relationship is broken at high damping, as can be seen at $\zeta_T\gtrsim10^{-2}$, due to the loss of the transition behavior, as was observed in Fig.~\ref{KEratio_mv_zetaN_sweeping}(a). Revisiting the question of the effect of impactor conditions on $N_{transition}$, it appears that rather than $N_{transition}$ simply increasing with impactor velocity, as might be assumed from Fig.~\ref{KEratio_transition}(a), there is a crossing of the $N_{transition}$ vs. $\zeta_T$ lines, such that at $\zeta_T \lesssim 10^{-3}$, $N_{transition}$ increases with velocity, with the opposite trend at $\zeta_T \gtrsim 10^{-3}$. However, we also note that this is close to the damping level above which the transition phenomena ceases. 

\begin{figure}[h!]
\centering
  \includegraphics[width=6.8in]{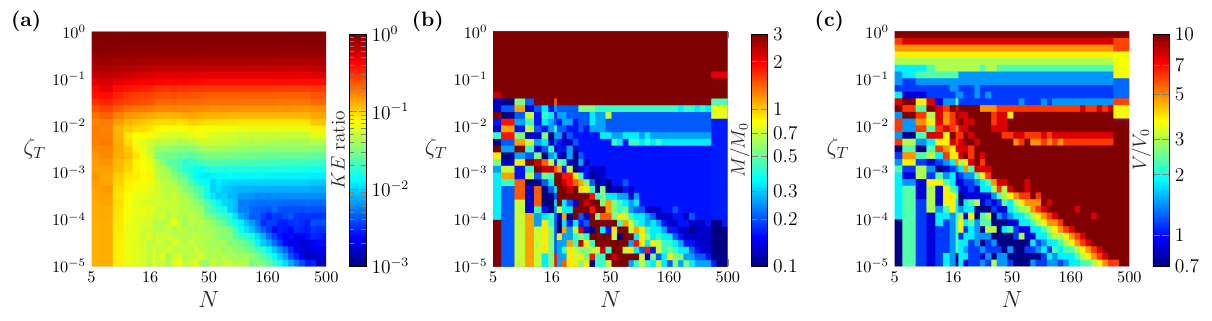}%
\caption{\textbf{(a) ``Best-of-the-best'' KE ratio as a function of nondimensional damping ratio $\bm{\zeta_T}$ and discreteness $\bm{N}$ picking from a sweep of impact conditions} ($M/M_0=10^{-1}$ to $10^{1}$ and $V/V_0=10^{-1}$ to $10^{1}$ using an exponential interval of 0.1). (b) Impactor mass ($M/M_0$) and (c) velocity ($V/V_0$) associated with the KE ratio denoted in (a).}\label{KEratio_mv_zetaN}
\end{figure}

Along the lines of impactor dependence, we also sweep a range of impactor conditions ($M/M_0=10^{-1}$ to $10^{1}$ and $V/V_0=10^{-1}$ to $10^{1}$) for the same range of  $N$ and $\zeta_T$ as in Fig.~\ref{KEratio_mv_zetaN_sweeping}(a), and measure the KE ratio. In Fig.~\ref{KEratio_mv_zetaN}, we plot the minimum (best) KE ratio (panel a), and their associated impact conditions, for each $N$ and $\zeta_T$ (mass and velocity in panels b and c, respectively). The result of this ``best-of-the-best'' parameter variation is KE ratios down to a minimum of $1.4\times10^{-3}$ (nearly three orders of magnitude improvement in kinetic energy transmission over the linear system) at $\zeta_T=2.585\times10^{-5}$, $N=330$, $M/M_0=0.1622$, and $V/V_0=10$. The general trend observed is that the bistable energy absorption mechanism relative performance (compared to linear materials) is best at low impactor masses $M/M_0~\sim10^{-1}$ (as can be seen by the blue triangular region in Fig.~\ref{KEratio_mv_zetaN}(b)) and high impactor velocities $V/V_0~\gtrsim10^{1}$ (as can be seen by the analogous red triangular region in Fig.~\ref{KEratio_mv_zetaN}(c)). In addition, considering KE ratio for the ``best-of-the-best'' cases, the same general phenomena as described above can be seen, where better KE ratio is associated with lower damping given that a threshold of sufficiently fine discretization is met (a threshold that increases as damping decreases). This general phenomena is preserved, albeit the ideal impact condition for which this best-of-the-best performance occurs shows significant variability around the previously described transition region (and region of moderate performance improvement by the bistable system). Comparing the observed best KE ratio in Fig.~\ref{KEratio_mv_zetaN} to those seen in Fig.~\ref{Fig4combo}, which are lower performing (higher KE ratio), we note that the middle row of Fig.~\ref{Fig4combo}, corresponding to the lower damping of $\zeta_T=4\times10^{-5}$, does not reach sufficiently fine discretization ($N=330$) in Fig.~\ref{Fig4combo} to reach its best performance. 

\section{An explanation for performance increase with increased discretization in the presence of nonlinearity and damping}\label{damping_layers}

\begin{figure}[h!]
\centering
  \includegraphics[width=5in]{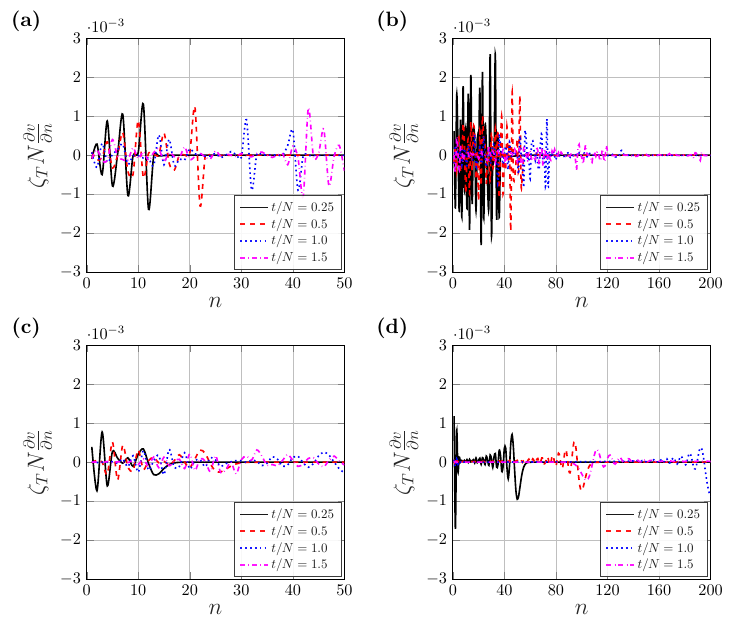}%
\caption{\textbf{Demonstration of the fourth key finding}. Normalized viscous force of (a,b) bistable metamaterial and (c,d) linear material with respect to unit cell index for an impact condition of $M/M_0=2$ and $V/V_0=2$ and bulk damping of $\zeta_T=4\times10^{-5}$. (a,c) $N=50$, (b,d) $N=200$. The fourth key finding is that the dependence on discretization stems from the partition of energy into trains of solitary waves that have pulse lengths proportional to the unit cell size, where, with intersite viscosity, the solitary wave trains induce high velocity gradients and thus enhanced damping compared to linear, and low-unit-cell-number bistable, materials.}\label{diff_velocity_m2v2}
\end{figure}

\noindent To explore why the relative performance of the damped bistable metamaterial is so much better than the undamped bistable system and the linear system, particularly at fine discretization (high $N$), we revisit impact conditions corresponding to Fig.~\ref{schematic_bistable_metamateirl}(d) (which are also the same as Fig.~\ref{KEratio_mv_zetaN_sweeping}, Fig.~\ref{KEratio_transition}, and Fig.~\ref{SI_2_KE_m2v1eta_m2v2eta}(b,d)). Since the highest relative performance arises when a small amount of damping is present, and our model uses an intersite linear viscosity, we plot in Fig.~\ref{diff_velocity_m2v2} the spatial profile of the viscous force at several times for both the damped bistable and linear systems. We specifically study the linear and bistable material (top and bottom row of Fig.~\ref{diff_velocity_m2v2}, respectively) at $\zeta_T=4\times10^{-5}$. This corresponds to the black and dashed-black lines in Fig.~\ref{schematic_bistable_metamateirl}(d) at $N$ values between which the performance of the linear and nonlinear systems switches. The left column Fig.~\ref{diff_velocity_m2v2} corresponds to $N=50$ and the right column Fig.~\ref{diff_velocity_m2v2}(b,d) corresponds to $N=200$. What can be seen in Fig.~\ref{diff_velocity_m2v2} is for the bistable cases, there are more, sharper spikes (solitary waves) of higher viscous force for $N=200$ than $N=50$, noting that sharper spikes are equivalent to higher frequency content. Conversely in the linear $N=200$ case, although there are higher frequency oscillations than in linear $N=50$ case, their viscous force is only substantially greater than the shorter discretization case near particle $n=0$. Comparing the bistable $N=200$ case to all the other three cases, we see the viscous force amplitude is higher, by over a factor of 2,  than all other cases, and the frequency content appears the highest. 

\begin{figure}[h!]
\centering
  \includegraphics[width=6in]{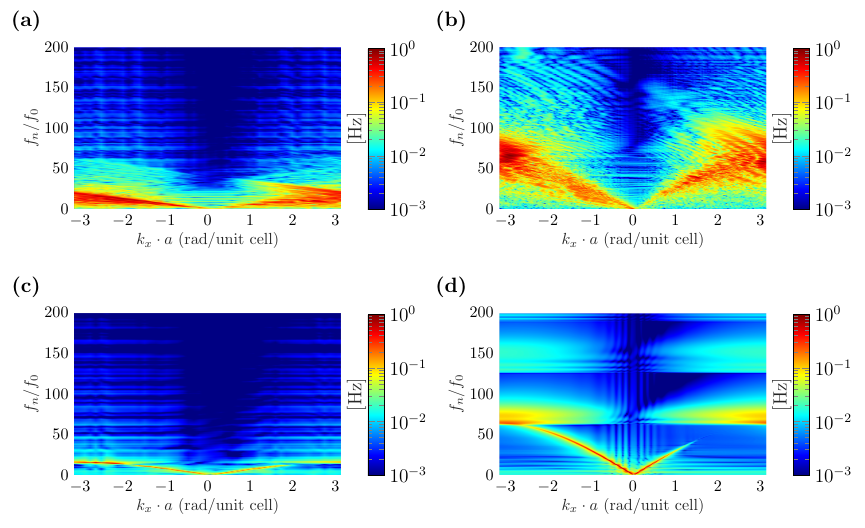}%
\caption{\textbf{Fourier transforms of the spatiotemporal normalized viscous force response of the bistable [top row (a,b)] and linear system [bottom row (c,d)] corresponding to the panels in Fig.~\ref{diff_velocity_m2v2}}. [Left column (a,c)] $N=50$, [right column (b,d)] $N=200$. Temporal frequency is $f_n/f_0$ and wave number is $k_x$, where $f_0=\sqrt{k_T/m_T}/2\pi$. }\label{comboFFTs}
\end{figure}

To confirm the relative frequency content in the four cases, in Fig.~\ref{comboFFTs} we plot the 2D Fast Fourier Transform (FFT) of the spatiotemporal viscous force response, corresponding to the four cases in Fig.~\ref{diff_velocity_m2v2}. We expect the same linearized dispersion for both linear and bistable systems, due to the same mass and linear stiffness. The upper cutoff frequency of the acoustic branch however is proportional to $N$, such that the cutoff frequency of the left column ($N=50$) of Fig.~\ref{comboFFTs} is four times smaller than the right column ($N=200$). This is because the total system linear stiffness and mass remains fixed, such that the intersite linear stiffness increases proportionally to $N$ and the layer mass decreases proportionally to $N$. An important distinction can thus be seen between the high $N$ bistable system response (Fig. \ref{comboFFTs}(b)) and the other three cases, where there is more energy (large red region) at higher frequencies. This higher frequency is allowed by the frequency conversion enabled by the system nonlinearity, coupled with the higher cutoff frequency of the linearized spectrum, allowing better phase matching at higher frequencies. 

\begin{figure}[h!]
\centering
  \includegraphics[width=4in]{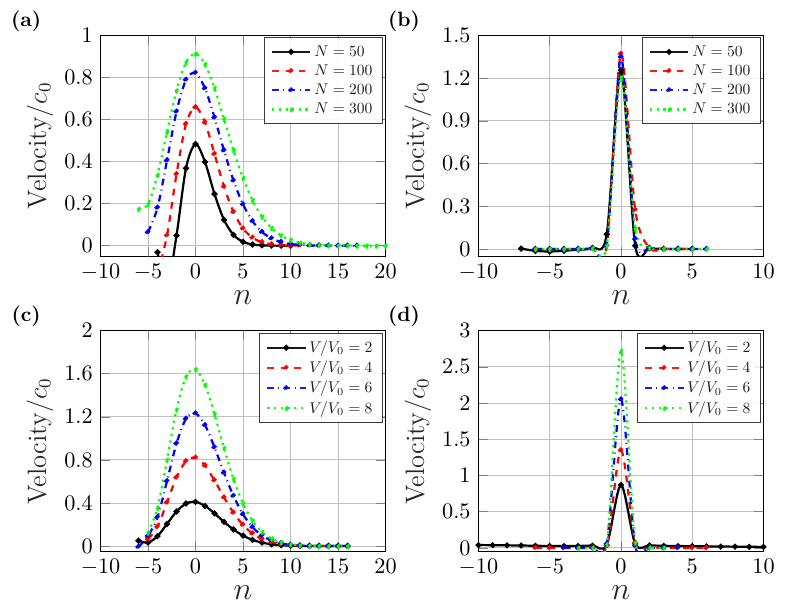}%
\caption{\textbf{Spatial profiles of solitary waves at time of $t/N=0.5$ in the linear [left column (a,c)] and bistable [right column (b,d)] systems for fixed impact conditions ($\bm{M/M_0=0.5}$ and $\bm{V/V_0=4}$) with varying discreteness [top row (a,b)], and  fixed discreteness ($\bm{N=200}$) with different impactor velocities [bottom row (c,d)].}}\label{soliton_N}
\end{figure}

Now we consider what nonlinear process enables the described frequency conversion. Given the aforementioned observed solitary wave phenomena existing in the bistable chain (\textit{e.g.} Fig.~\ref{KE_m05v4_N_undamped}(d-f) and Fig.~\ref{KE_m05v4_N_damped}(d-f)), we suggest that the self-localization involved in solitary wave formation is the underlying mechanism driving the frequency conversion. For many types of solitary waves, the pulse length is tied to a number of unit cells \cite{nesterenko2013dynamics,katz2019solitary}, making the pulse length proportional to the unit cell size. In Fig.~\ref{soliton_N}, we study the effect of discretization ($N$) and impactor velocity ($V$) on the pulse shape for both the linear and bistable system. In the linear material the pulse is selected by identifying the position of peak velocity amplitude at time $t/N=0.5$, which we denote $n=0$ in Fig.~\ref{soliton_N}. In the bistable material, the position of the peak of the first solitary wave within the generated solitary wave train is chosen, and similarly set to $n=0$. The left column of Fig.~\ref{soliton_N} corresponds to the linear material and the right column of Fig.~\ref{soliton_N} corresponds to the nonlinear material. The top row of Fig.~\ref{soliton_N} shows the effect of changing discretization (all other parameters kept constant). It can be seen for the linear system (Fig.~\ref{soliton_N}(a)) that the pulse width encompasses more particles with finer discretization (similar physical pulse length). In contrast, as can be seen in Fig.~\ref{soliton_N}(b), the solitary waves in the bistable material retain a fixed number of particles for their width, resulting in physical solitary wavelengths proportional to the unit cell size. The bottom row of Fig.~\ref{soliton_N} shows the effect of changing impactor velocity (all other parameters kept constant), where it can be seen that the pulse length does not change with increased impactor velocity, however the amplitude of the pulse increases approximately proportionally. This latter point is important, as given the fixed pulse length but increasing amplitude, this means that higher impactor velocities leads to higher spatial velocity gradients, or equivalently, higher viscous forces. This explains the higher KE ratio seen in Fig.~\ref{KEratio_transition}(a) for the light mass, high velocity impactor (dashed grey line), as well as the dark blue regions in Fig. \ref{Fig4combo}(h,k,l) and the high impact velocities identified as the ``best-of-the-best'' in Fig.~\ref{KEratio_mv_zetaN}. However, the observation of increasing viscous force at higher impact velocities does not mean that higher impactor velocities alone result in higher KE ratio, as it must be accompanied by lower impactor mass, so that the material is not effectively ``overpowered'' by excess impactor kinetic energy, as can be seen in the top right regions of all panels in Fig.~\ref{Fig4combo}. Another important point is that this highlights that there are two mechanisms contributing to mitigating the propagation of kinetic energy, namely, the elastic trapping of energy via bistability and the enhanced dissipation via the interplay of nonlinear self-localization and viscosity. These two different mechanisms can be observed in Fig.~\ref{SI_2_KE_m2v1eta_m2v2eta}, corresponding to panels (a) and (b), respectively.

\begin{figure}[h!]
\centering
 \includegraphics[width=2.75in]{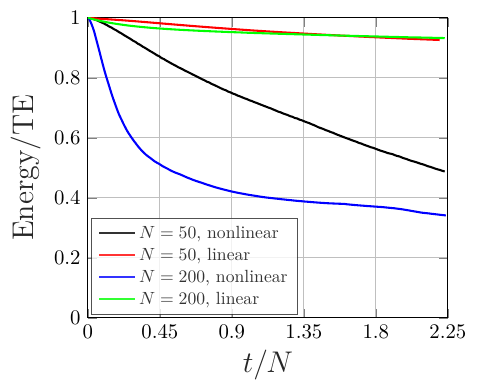}%
\caption{\textbf{Total energy of the system (as a function of time) normalized by initial total energy (initial impactor kinetic energy) versus normalized time for the bistable metamaterial and linear materials}. Bulk damping of $\zeta_T=4\times10^{-5}$ experiencing an impact condition of $M/M_0=2$ and $V/V_0=2$.}\label{Total_Energy_M2V2}
\end{figure}

\section{Dimensional analysis}\label{dimensional}

\noindent To further explore the system parameters and physical mechanisms underlying this observed phenomena of discretization and damping interplay combined with impactor condition dependence, we use the Buckingham-Pi algorithm \cite{buckingham1914physically} to define dimensionless parameters for our linear comparative system. Because there are $n_\pi=7$ dimensional variables characterizing the system ($m_{im}$, $V^\ast$, $m$, $c_{01}^\ast$, $\eta_T$, $a$, and $L$), which have $k_\pi=3$ independent base units (kg, m, and s), we can define $p_\pi=n_\pi-k_\pi=4$ non-dimensional parameters to characterize our system, as follows: 
\begin{equation}\label{pi_dimensionless parameters}
\begin{split}
    & \Pi_M=M, \\
    & \Pi_N=N, \\
  &  \Pi_{I-E}=\frac{m_{im} V^{\ast2}}{c_{01}^\ast L^2}=\frac{Mm_T (Vc_0)^2}{k_T L^2}=MV^2,\\
  &  \Pi_{I-V}=\frac{m_{im} V^\ast}{L \eta_T}=\frac{Mm_T (Vc_0)}{L\zeta_T 2\sqrt{m_Tk_T}}=\frac{MV}{2\zeta_T},
\end{split}
\end{equation}
where $\Pi_{I-E}$ represents the relationship between the impactor's inertial (kinetic) energy and the linear sample's elastic energy when it is fully compressed, and $\Pi_{I-V}$ signifies the connection between and the amount of viscous momentum required to flow across the sample and the impactor's (inertial) momentum.

\begin{figure}[h!]
\centering\includegraphics[width=6.8in]{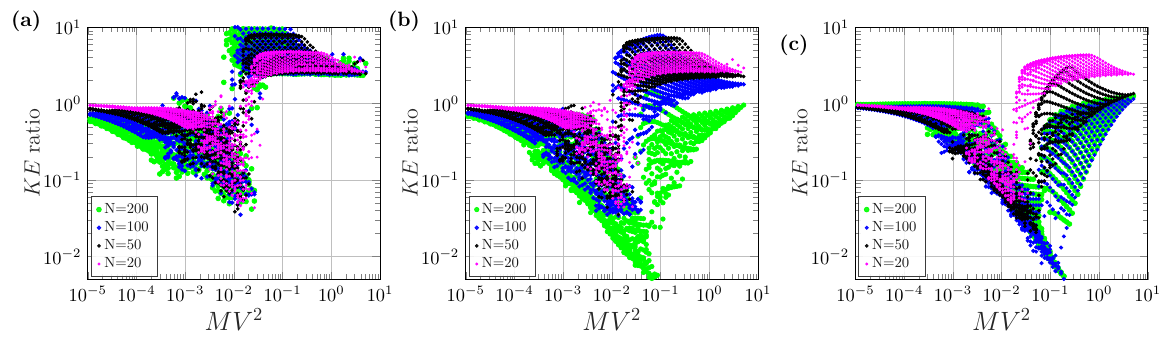}%
\caption{\textbf{Demonstration of the fifth key finding, and KE ratio vs. nondimensional ratio of impactor inertial to material elastic energy ($\bm{\Pi_{I-E}}$) for various bulk damping values ($\bm{\zeta_T}$) and numbers of unit cells ($\bm{N}$).} Impact conditions are varied across the range $M/M_0=10^{-1}$ to $10^{1}$ and $V/V_0=10^{-1}$ to $10^{1}$. (a) $\zeta_T=0$ (b) $\zeta_T=4\times10^{-5}$, and (c) $\zeta_T=4\times10^{-4}$. The fifth key finding is when sufficiently fine discretization has been reached at low viscosities, the bistable system outperforms the linear one for a wide range of impactor conditions.}\label{KE_mv2_zetaN}
\end{figure}

In Fig. \ref{KE_mv2_zetaN}, we study the dependence of the KE ratio on $\Pi_{I-E}$. We consider impactor parameters across the range $M/M_0=10^{-1}$ to $10^{1}$ and $V/V_0=10^{-1}$ to $10^{1}$, for three values of bulk damping ($\zeta_T=0$, $4\times10^{-5}$, and $4\times10^{-4}$, corresponding to panels (a), (b), and (c), respectively). For the undamped case of Fig.~\ref{KE_mv2_zetaN}(a), we see, as in the top row of Fig.~\ref{Fig4combo} that it is possible for the bistable material to outperform the linear material. There is however limited magnitude of improvement ($\sim 10^{-1}$ KE ratio) compared to the damped cases. Further, there is minimal effect of discretization ($N$) on the resulting performance, which is consistent with the lack of the underlying nonlinear self-localization and viscosity interplay mechanism. We also highlight that the peak performance in the undamped case occurs at  $\Pi_{I-E}\sim 10^{-2}$ which aligns well with the predicted nominal impact condition of $M_0 V_0^2=10^{-2}$. In contrast to the undamped case, in Fig. \ref{KE_mv2_zetaN}(b,c), we see that, as described previously throughout this manuscript, with the introduction of damping comes a dependence on minimum KE ratio with $N$, such that peak performance is reached after some critical threshold of $N$ and decreased $\zeta_T$ results in a higher critical $N$. Further, again as described herein, the minimum KE ratio (peak performance) in the damped cases are $\sim 10\times$ better than in the undamped case, due to the added nonlinear self-localization and viscosity interplay. We also note that the peak performance of the damped cases of Fig. \ref{KE_mv2_zetaN}(b,c), occurs at $\Pi_{I-E}\sim 10^{-1}$, which is consistent with the impactor conditions for peak performance from Fig.~\ref{KEratio_mv_zetaN} of $MV^2=10M_0V_0^2$. Finally, we observe that in both damped and undamped cases, this point of best performance is of further importance, as it aligns with an impact conditions threshold after which larger $MV^2$ contain scenarios where the performance of the bistable material is worse (or minimally better) than the linear system. However, there is a significant distinction in this regard between the damped and undamped cases. In the damped cases, at $MV^2$ larger than the conditions of best performance, the performance is significantly worse than the linear material. In contrast, in the damped cases, we see a similar behavior at low $N$, however, when sufficiently high $N$ is reached for best performance, the capacity for poor performance is limited, even with changing impactor conditions, as can be seen by the green dataset in Fig.~\ref{KE_mv2_zetaN}(b) and the green and blue datasets in Fig.~\ref{KE_mv2_zetaN}(c), corresponding to $\zeta_T=4\times10^{-5}$ and $\zeta_T=4\times10^{-4}$, respectively. This is the fifth key finding, namely, when sufficiently fine discretization has been reached at low viscosities, the bistable system outperforms the linear one for a wide range of impactor conditions. Considering 
this performance threshold from an applied system design perspective, there is a clear benefit to ensuring sufficiently high discretization is reached, in that it increases the potential performance and decreases the magnitude (or eliminates altogether) the possibility of worse performance that the linear system. On the other had, if a high sensitivity of kinetic energy transmission to impact conditions is desired, perhaps for applications other than impact mitigation, then a lower discreteness could be desirable. 

\begin{figure}[h!]
\centering
\includegraphics[width=5in]{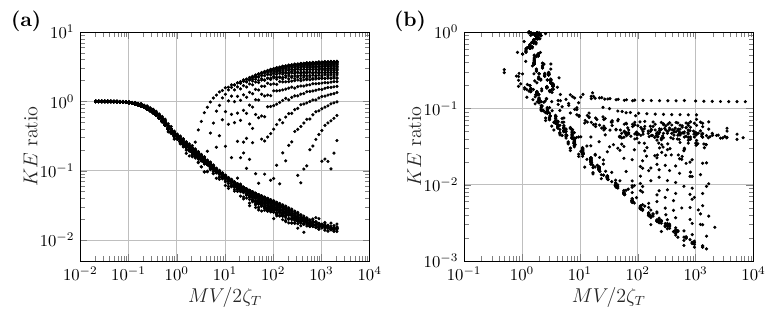}%
\caption{\textbf{KE ratio versus non-dimensional ratio of impactor momentum to viscous momentum required to flow across the sample corresponding to (a) the data of Fig. \ref{KEratio_mv_zetaN_sweeping}(a) for fixed impactor conditions, and to (b) the data of Fig. \ref{KEratio_mv_zetaN}(a) representing the ``best-of-the-best'' performance for varied impactor conditions at each discretization ($\bm{N}$) and damping ($\bm{\zeta_T}$).}}\label{KE_mv_zetaN}
\end{figure}

To incorporate the effects of damping into the nondimensional parameter study, we first note that the non-dimensional parameter of $N/\zeta_T$ plotted in Fig.~\ref{KEratio_mv_zetaN_sweeping}(d) is equivalent to a constant factor with $\Pi_{N}\Pi_{I-V}$ (\textit{i.e.}, a discretization weighted ratio of impactor to viscous momentum) when the impact conditions are kept constant (as they are in Fig.~\ref{KEratio_mv_zetaN_sweeping}). In Fig.~\ref{KE_mv_zetaN}(a), we plot $\Pi_{I-V}$ for the single impactor condition of Fig.~\ref{KEratio_mv_zetaN_sweeping}. This contrasts with Fig.~\ref{KEratio_mv_zetaN_sweeping}(d), in that $N$ is not divided out in Fig.~\ref{KE_mv_zetaN}(a). In Fig.~\ref{KE_mv_zetaN}(a), the high performing KE ratios collapse together more than in Fig.~\ref{KEratio_mv_zetaN_sweeping}(d). Considering low $\zeta_T$, two groupings can be observed, one with low and one with high KE ratio. Taking a horizontal slice (fixed $\zeta_T$) of Fig.~\ref{KEratio_mv_zetaN_sweeping}(a), one can see the groupings correspond to above or below the critical value of discretization $N$, wherein once sufficiently fine $N$ has been reached, there is minimal improvement in the KE ratio (also observable in Fig.~\ref{KEratio_mv_zetaN_sweeping}(b)). Such groupings let us conclude that while the $N_{threshold}$ and $\zeta_T$ are clearly correlated (as per Fig.~\ref{KEratio_transition}(b)), the post-threshold KE ratio (performance) achievable, within a fixed impact conditions and elastic nonlinearity setting, can be predicted by $\zeta_T$ without $N$, given that the critical threshold of discretization has been met. 

We next plot the broader ``best-of-the-best'' dataset from Fig.~\ref{KEratio_mv_zetaN} (sweeping over $M$ and $V$, picking the best KE ratio, and associating that with each value of $N$ and $\zeta_T$). Figure~\ref{KE_mv_zetaN}(b) shows the same quantity as Figure~\ref{KE_mv_zetaN}(a) ($\Pi_{I-V}$), but now the dataset includes varied impactor conditions. Because only the optimal cases are chosen (from a set of swept $M$ and $V$) for each $N$ and $\zeta_T$, unlike Fig.~\ref{KE_mv_zetaN}(a), the KE ratio does not exceed unity in Fig.~\ref{KE_mv_zetaN}(b,c). The ``front'' of minimum KE ratio can be seen analogously to Fig.~\ref{KE_mv_zetaN}(a), but now with approximately a further $10\times$ improvement in performance. The retention of this front for varied impactor conditions suggests that the optimum performance for fixed system elastic nonlinearity can be predicted via $\Pi_{I-V}$ (the balance of impactor to viscous momentum) given that a critical level of discretization has been met.

\section{Conclusion}\label{conclusion}

\noindent In this work, apart from high sensitivity to impactor conditions \cite{fancher2023dependence}, we show that the kinetic energy transmission performance of a simulated bistable mechanical metamaterial also significantly depends on the interplay of viscous damping and the discretization of the material. Specifically, we make the following five key findings: 1) The bistable material can counter-intuitively perform better (to nearly $10^3 \times$ better than the linear system) as the viscosity \emph{decreases} (but remains finite), but only when sufficiently fine discretization has been reached (\textit{i.e.} the system approaches sufficiently close to the continuum limit); 2) This discretization threshold is sharp, and depends on the viscosity present; 3) The bistable materials can also perform significantly worse than linear systems (for low discretization and viscosity or zero viscosity); 4) The dependence on discretization stems from the partition of energy into trains of solitary waves that have pulse lengths proportional to the unit cell size, where, with intersite viscosity, the solitary wave trains induce high velocity gradients and thus enhanced damping compared to linear, and low-unit-cell-number bistable, materials; and 5) When sufficiently fine discretization has been reached at low viscosities, the bistable system outperforms the linear one for a wide range of impactor conditions. The first point introduces the existence of a nonlinear dynamical ``size effect''. This nonlinear self-localization and viscous size effect is accompanied by a change in optimal impactor conditions from the undamped case, such that the system becomes particularly suited for very light mass ($1.59\%$ of the system mass) and high speed ($260\%$ of the lattice sound speed) impactors. Such high speed impactors suggest the future study of high strain rate effects, and irreversible phenomena such as plasticity \cite{wallen2024strongly} and fracture will be important. 

In addition, we find that the ratio of impactor to viscous momentum can serve as a strong predictor of optimal system performance, given a particular elastic nonlinearity and that a critical level of discretization has been reached. Further, this critical level of transition can also be well predicted by damping. Future investigations may explore related questions raised herein, \textit{e.g.}, how the particular form of the elastic nonlinearity quantitatively dictates the relationship between $N_{transition}$ and $\zeta_T$ as well as the KE ratio (performance) dependence on $\Pi_{I-V}$. We suggest that these bistable metamaterials may find particular use in applications where materials with properties characteristic of highly damped, viscous materials are undesirable. For instance, most viscous materials have relatively low modulus and maximum usable temperatures \cite{AshbyM.F1999Msim}. As such, these structured materials may be particularly advantageous where low damping is desired (\textit{e.g.}, when wanting low amplitude acoustic signals to be able to traverse the material) or in extreme, high temperature environments where normal high damping materials melt, yet impact mitigating properties are still desired. We note that the viscosity range considered herein corresponds to constituent materials such as higher damping ceramics and low damping metals, however we expect the phenomena is extensible to even lower damping ceramics \cite{lakes2009viscoelastic}. Given the large deformations typically involved in bistable mechanisms, we suggest that high strength ceramics \cite{bauer2019additive} may be of particular value in this regard. Future possibilities connected to the phenomena shown herein also include extension to: figures of merit for impact mitigation other than KE ratio; higher dimensions; other application areas involving bistability and waves, such as mechanical signal processing and shape changing; as well as investigation into the possible existence of similar viscosity-discreteness interplay in other nonlinear, but non-bistable, media.

\section*{Acknowledgements}
\noindent This project was supported by the Army Research Office and the Army Research Directorate of the DEVCOM Army Research Laboratory under grant no. W911NF-17-1-0595, and by the UC National Laboratory Fees Research Program of the University of California, Grant Number L22CR4520. I.F. acknowledges support from the Department of Defense (DoD) through the National Defense Science \& Engineering Graduate (NDSEG) Fellowship Program.
\section*{Conflict of Interest}
\noindent The authors declare no conflicts of interest.

\setlength{\itemsep}{1.2mm}\footnotesize
\bibliography{Damping_reference}


\begin{appendices}
\appendix
\section{Force displacement relationship of a nonlinear material for various discreteness}\label{Appendix_A}
\setcounter{figure}{0}
\renewcommand{\thefigure}{A.\arabic{figure}}

\begin{figure}[h!]
\centering
\subfloat[]{%
  \includegraphics[height=2.2in]{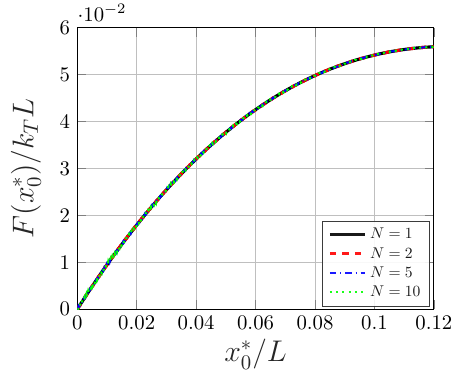}%
}
\subfloat[]{%
  \includegraphics[height=2.2in]{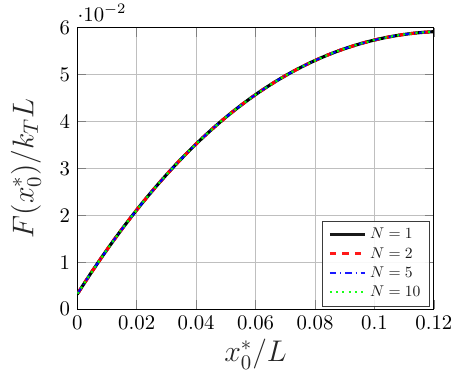}%
}
\caption{\textbf{Normalized quasi-static force-displacement responses (using a slow moving boundary condition with constant velocity) of a nonlinear material under various levels of discreteness ($\bm{N}$).} (a) Undamped and (b) with intersite damping.}\label{force_disp_N}
\end{figure}

\noindent To validate the uniform quasi-static response of our DEM with changing $N$, we apply a slow moving applied displacement (constant velocity) to the impactor particle. The resulting force-displacement relationship is shown in Fig.~\ref{force_disp_N}, with (panel a) and without (panel b) damping. Due to the presence of the damping, the expected upshift can be seen in  Fig.~\ref{force_disp_N}(b) compared to Fig.~\ref{force_disp_N}(a). The observed response is uniform for all values of $N$.  

\end{appendices}

\end{document}